\documentclass[]{article}
\newcommand{\nd}{\noindent}
\usepackage{graphics}
\usepackage{graphicx}
\usepackage{amsmath}
\usepackage{amssymb}
\usepackage{amstext}
\usepackage{amscd}
\usepackage{amsfonts}
\usepackage{xcolor}

\title{Thermodynamically consistent entropic-force cosmology}
\author{\small{D. J. Zamora$^{1,2}$\thanks{E-mail: javierzamora055@gmail.com}, C. Tsallis$^{1,3,4}$\thanks{E-mail: tsallis@cbpf.br}}, \\
\small{$^1$ Centro Brasileiro de Pesquisas Fisicas and}\\
\small{ National Institute of Science and Technology for Complex Systems,}\\
\small{ Rua Dr. Xavier Sigaud 150, Rio de Janeiro, 22290-180, Brazil.}\\
\small{$^2$ Instituto de Fisica del Noroeste Argentino,}\\
\small{ Av. Independencia 1800, Tucuman, CP 4000, Argentina}\\
\small{$^3$ Santa Fe Institute, 1399 Hyde Park Road, Santa Fe, 87501, NM, United States}\\
\small{(IFLP-CCT-CONICET)-C. C. 727, 1900 La Plata -
Argentina}\\
\small{$^4$  Complexity Science Hub Vienna, Josefstadter Strasse 39, Vienna, 1080, Austria}}

\date{\today}

\begin{document}

\maketitle

\begin{abstract}
We analyze the thermodynamical consistency of entropic-force cosmological models. Our analysis is based on a generalized entropy scaling with an arbitrary power of the Hubble radius. The Bekenstein-Hawking entropy, proportional to the area, and the nonadditive  $S_{\delta=3/2}$-entropy, proportional to the volume, are particular cases. One of the points to be solved by entropic-force cosmology for being taken as a serious alternative to mainstream cosmology is to provide a physical principle that points out what entropy and temperature have to be used. We determine the temperature of the universe horizon by requiring that the Legendre structure of thermodynamics is preserved. We compare the performance of thermodynamically consistent entropic-force models with regard to the available supernovae data by providing appropriate constraints for optimizing alternative entropies and temperatures of the Hubble screen. Our results point out that the temperature differs from the Hawking one.
\end{abstract}


\section{Introduction}
The lambda cold dark matter model ($\Lambda$CDM) assumes a cosmological constant $\Lambda$ and the existence of dark energy. This model is the simplest one that can explain an accelerated expansion of the late universe. However, it implies several theoretical peculiarities, such as the cosmic coincidence and the cosmological constant problem \cite{Weinberg1989,Carroll2001}. In order to handle these difficulties, several alternative models have been proposed \cite{Weinberg2008,Ellis2012,Komatsu2014,Sola2013}. 

An interesting model based on the concept of entropic-force is able to explain the accelerated expansion of the universe \cite{Easson2011,Easson2012}. From this standpoint, the controversial dark energy component is not necessary. An entropic-force is an emergent phenomenon resulting from the tendency of a thermodynamical system to extremize its entropy, rather than from a particular underlying fundamental force. There is no field associated with an entropic-force. The force equation is expressed in terms of spatial dependence of the entropy $S$. At a fixed temperature $T$, the entropic-force $F$, is given by

\begin{equation}
F=-T\frac{dS}{dr},
\end{equation}

\nd where $r$ is the radius of a cavity, assumed nearly isotropic.

At this point, let us make an important clarification. The present entropic-force cosmological model is definitively different from the idea that gravity itself is an entropic-force, as suggested in \cite{Verlinde2011}.

The Hubble sphere is a region of the observable universe beyond which, due to the expansion of the universe, objects appear to recede from the observer at a rate larger than the speed of light. Its radius is known as the Hubble radius $r_H$. Entropic-force models of cosmology are based on considering the surface of the Hubble sphere (Hubble surface or horizon) as a screen whose entropy and corresponding temperature are analogous to those of the horizon of a black-hole \cite{Easson2011}. That is, the Hubble surface would coincide with a cosmological event horizon (a boundary separating events that are visible at some time from those that are never visible). In the present paper, we use the Hubble surface as the screen since it coincides with the apparent horizon in a spatially flat universe \cite{Easson2011}. Entropic-force models lead to an extra driving term with regard to the so-called Friedmann equations \cite{Easson2011}. The entropic-force term has the potential of explaining the accelerated expansion without introducing new fields nor dark energy.

The first entropic-force model \cite{Easson2011} assumes that the entropy and temperature associated to the horizon of the universe are the Bekenstein-Hawking entropy \cite{Bekenstein1973} and the Hawking temperature \cite{Hawking1974}, respectively. After that, other entropies were considered, such as the nonadditive $S_{\delta=3/2}$-entropy \cite{Komatsu2013}. This entropy was proposed in \cite{Tsallis2013} in the context of black-holes. Let us remind the reader that additive Bekenstein-Hawking entropy is proportional to the area, whereas the nonadditive $S_{\delta=3/2}$-entropy is proportional to the volume (at least in the case of equal probabilities). The expression of the temperature of the Hubble horizon is currently not obtained from an neat physical principle \cite{Tu2018}. However, it is usually assumed to be the Hawking temperature expressed in terms of the universe parameters, namely

\begin{equation}
T_{BH}=\frac{\hbar c}{2\pi k_Br_H}=\frac{\hbar H}{2\pi k_B},
\label{Thawking}
\end{equation}

\nd where $c$ is the speed of light, $\hbar$ is the reduced Planck constant, $k_B$ the Boltzmann constant, and $H=H(t)$ is the Hubble parameter. $H$ is defined as 

\begin{equation}
H\equiv\frac{c}{r_H}=\frac{\dot{a}}{a},
\end{equation}

\nd $a=a(t)$ being the scale factor, a dimensionless quantity  parametrizing the relative expansion of the universe.

In mainstream cosmology, matter and space-time emerged from a singularity and evolved through four distinct periods, namely, early inflation, radiation, dark matter, and late-time inflation (driven by dark energy according to the $\Lambda$CDM model). During the radiation and dark matter dominated stages, the universe is decelerating while the early and late-time inflation are accelerating stages. A possible connection between the accelerating periods remains unknown, and, even more intriguing, the most popular dark energy candidate powering the present accelerating stage ($\Lambda$-vacuum) relies on the cosmological constant and coincidence puzzles.

The entropic-force term is to be added within the acceleration and continuity Friedmann equations \cite{Easson2011,Easson2012}. This extra term depends on $H^2$ and affects the background evolution of the universe. We do not focus here on the inflation of the early universe. It has been shown that entropic-force models which include $H^2$ terms are not able to describe on a single footing both decelerating and accelerating stages \cite{Perico2013,Komatsu2013b}. Basilakos et al. \cite{Basilakos2012} have shown that the first Easson-Frampton-Smoot (EFS) entropic-force model (which includes the $H^2$ term) does not describe properly both acceleration and deceleration cosmological regimes unless a $\dot H$ term is included.

This motivated the use of alternative entropic measures. Komatsu and Kimura (KK) proposed a modified entropic-force model \cite{Komatsu2013} using the $S_{\delta=3/2}$ entropy. In this class of models, the extra entropic-force terms depend on the class of entropy being used. For example, $H^2$ terms are derived from an area-scaling entropy \cite{Easson2011}, whereas $H$ terms are derived from a volume-scaling entropy \cite{Komatsu2013}. A modified entropic-force model which includes $H$ terms is capable of describing both decelerating and accelerating regimes. Moreover, it has been argued that bulk viscous models (which include $H$ terms) are hard to reconcile with astronomical observations of structure formations \cite{Li2009}. This suggests that it is necessary to consider not only an $H$ term but also a constant entropic-force term. 

It turns out, however, that some of these entropic-force models violate the Legendre structure of thermodynamics, as will became clear below. Our present aim is to point out that the entropy and temperature of the Hubble horizon cannot be freely chosen. These quantities are related by the Legendre structure and the use of a modified entropy introduces a corresponding modification in the temperature. We will show that, whatever scaling entropy one uses, thermodynamically consistent entropic-force models yield an $H^2$ extra term. Consequently, all entropic-force models using a single entropy are unable to explain both accelerating and decelerating regimes.

%911+16x3=959

\section{Thermodynamical entropy}
After the works of Bekenstein \cite{Bekenstein1973} and Hawking \cite{Hawking1974}, it is common in the literature to accept that the black-hole entropy violates thermodynamical extensivity, meaning that the entropy of a $d=3$ black-hole is proportional to the area of its boundary instead of being proportional to its volume. To recover thermodynamical extensivity, $S_\delta$ was introduced \cite{Tsallis2013}. This nonadditive entropy is defined as

\begin{equation}
S_\delta=k_B\sum_i^Wp_i\left(\ln\frac{1}{p_i}\right)^\delta,\;(\delta>0),
\end{equation}

\nd where $\delta=1$ recovers the Boltzmann-Gibbs entropy. For equal probabilities, 

\begin{equation}
\frac{S_{\delta=d/(d-1)}}{k_B}\propto\left(\frac{S_{BH}}{k_B}\right)^{d/(d-1)},
\end{equation}

\nd where $d$ is the dimension, being thus connected with the well known Bekenstein-Hawking entropy $S_{BH}$. For $d=3$ we have $S_{\delta=3/2}/k_B\propto\left(S_{BH}/k_B\right)^{3/2}$.

Let us focus now on the free energy $G$ of a generic d-dimensional system

\begin{equation}
\begin{split}
G(V,T, p,\mu,...)&=U(V,T,p,\mu,...)-TS(V,T,p,\mu,...)\\
&+pV-\mu N(V,T,p,\mu,...)-...,
\end{split}
\label{Legendre}
\end{equation}

\nd where $T,p,\mu$ are the temperature, pressure, and chemical potential, and $U,S,V,N$ are the internal energy, entropy, volume, and the number of particles, respectively. One can distinguish in this Legendre transformation three different types of variables (see \cite{Tsallis2013} and references therein), namely (i) those that are expected to always be extensive ($S,V,N,...$), i.e., scaling with $V=L^d$, where $L$ is a characteristic linear dimension of the $d$-dimensional system, (ii) those characterizing the external conditions under which the system is placed ($T, p,\mu,...$), scaling with $L^\theta$, and (iii) those representing energies ($G,U$), scaling with $L^\varepsilon$. From Eq. (\ref{Legendre}), it trivially follows

\begin{equation}
\varepsilon=\theta+d,
\label{key}
\end{equation}

\nd where standard thermodynamics (short-range interactions) corresponds to $\theta=0$.
Let us now consider a Schwarzschild (3+1)-dimensional black hole. In this case, the energy scales like the mass $M_{bh}$, which in turn scales with $L$ \cite{Weinberg1972,Hughston1990,Taylor2000}. Therefore, $\varepsilon=1$, hence,
\begin{equation}
\theta=1-d.
\label{key2}
\end{equation}

If we physically identify the black hole with its event horizon surface, then it has to be considered as a $d=2$ system, then $\theta=-1$, which recovers the usual Bekenstein-Hawking (BH) scaling $T\propto1/L\propto1/M_{bh}$, Eq. (\ref{Thawking}). However, if the black hole is to be considered as a $d=3$ system, which legitimates using the $S_{\delta=3/2}$ entropy, hence $\theta=-2$, i.e., $T$ scales like $1/L^2\propto1/M_{bh}^2$.

This is a crucial point since, unless we would be willing -- which is not our case -- to violate Eq. (\ref{Legendre}),
the Hawking temperature can not be the temperature to be used in a cosmological model if the chosen entropy does not scale with the area. The simultaneous use, for a black hole, of the Hawking temperature and of an entropy differing from the Bekenstein-Hawking one leads to a violation of the thermodynamical Legendre structure. When working with an entropic-force cosmological model based on entropies different from the Bekenstein-Hawking one, there are two options, (i) to preserve the Hawking temperature for the horizon, which contradicts thermodynamics, or (ii) to work with a model consistent with thermodynamics by modifying the temperature. We consider thermodynamics one of the most fundamental physical theories, and therefore we only explore the second option here.

%430+16x7=430+112=542
%+959=1501 en total

\section{Cosmological models for $L^d$-scaling entropies}

Our present goal is to study entropic-force cosmology consistently with thermodynamics. In order to make a general discussion, let us consider an entropy that scales with length with some arbitrary positive power $d\in\mathbb{R^+}$. This includes the Bekenstein-Hawking entropy (d=2) and the $\delta=3/2$ entropy (d=3) as particular cases.

Since the Planck length is $L_P=\sqrt{\hbar G/c^3}$, the Bekenstein-Hawking entropy can be expressed as

\begin{equation}
S_{BH}=k_B\pi\left(\frac{r_H}{L_P}\right)^2,
\label{sbh}
\end{equation}

\nd and the Hawking temperature as

\begin{equation}
T=\frac{T_P}{2\pi}\frac{L_P}{r_H},
\end{equation}

\nd where $T_P=\sqrt{\hbar c^5/G k_B^2}$ is the Planck temperature. Let us suppose then, a generalized entropy of the form

\begin{equation}
S=k_BA_d\left(\frac{r_H}{L_P}\right)^d,
\label{Snu}
\end{equation}

\nd where $A_d$ is a dimensionless factor. According to Eq. (\ref{key2}), the thermodynamically correct temperature must scale like $T\propto r_H^{1-d}$. Consequently, we propose

\begin{equation}
T=\frac{T_P}{B_d}\left(\frac{r_H}{L_P}\right)^{1-d},
\end{equation}

\nd where $B_d$ is a dimensionless factor. Therefore, We can see that the EFS entropic-force model \cite{Easson2011} is consistent with thermodynamics, whereas the KK model introduced in \cite{Komatsu2013} is not. This is so because the Hawking temperature is the corresponding temperature for the Bekenstein-Hawking entropy, but not for the $\delta=3/2$ entropy. Naturally, entropies differing from those can be used in entropic-force cosmological models. In what follows, we study a rather generic thermodynamically consistent entropic-force model. The entropic force is given by

\begin{equation}
F\equiv-T\frac{dS}{dr_H}=-k_B\frac{d\, A_d}{B_d}.\frac{T_P}{L_P}\equiv-C_dF_P,
\end{equation}

\nd where $F_P\equiv k_BT_P/L_P=c^4/G$ is the Planck force, and $C_d\equiv d\,A_d/B_d$. Therefore, the entropic pressure in the Hubble surface is

\begin{equation}
p_F\equiv\frac{F}{4\pi r_H^2}=-\frac{C_dF_P}{4\pi r_H^2}=-\frac{C_d\,c^2}{4\pi G}H^2.
\end{equation}

This pressure is precisely $C_d$ times the entropic pressure calculated in the EFS model \cite{Easson2011}, which we recover for $C_d=1$. To obtain the Friedmann equations modified by $p_F$, we replace the effective pressure $p'=p+p_F$ in the acceleration equation

\begin{equation}
\frac{\ddot{a}}{a}=-\frac{4\pi G}{3}\left(\rho+\frac{3p'}{c^2}\right),
\label{acc}
\end{equation}

\nd thus arriving to

\begin{equation}
\frac{\ddot{a}}{a}=-\frac{4\pi G}{3}\left(\rho+\frac{3p}{c^2}\right)+C_dH^2.
\label{accelerationeq}
\end{equation}

In Eq. (\ref{acc}) and (\ref{accelerationeq}), $\rho$ is the total energy density of the universe. Replacing now $p'$ in the continuity equation 

\begin{equation}
\dot{\rho}+3\frac{\dot{a}}{a}\left(\rho+\frac{p'}{c^2}\right)=0,
\end{equation}

\nd we obtain

\begin{equation}
\dot{\rho}+3\frac{\dot{a}}{a}\left(\rho+\frac{p}{c^2}\right)=\frac{3C_d}{4\pi G}H^3.
\label{continuityeq}
\end{equation}

Now, we follow the procedure in \cite{Komatsu2013} to derive a modified Friedmann equation from Eqs. (\ref{accelerationeq}) and (\ref{continuityeq}), since only two of the three are independent. The generalized Friedmann and acceleration equations

\begin{equation}
\left(\frac{\dot{a}}{a}\right)^2=\frac{8\pi G\rho}{3}+f(t),
\end{equation}
\begin{equation}
\frac{\ddot{a}}{a}=-\frac{4\pi G}{3}\left(\rho+\frac{3p}{c^2}\right)+g(t),
\label{generalizedacceleration}
\end{equation}

\nd imply

\begin{equation}
\dot{\rho}+3\frac{\dot{a}}{a}\left(\rho+\frac{p}{c^2}\right)=\frac{3}{4\pi G}H\left(-f(t)-\frac{\dot{f}(t)}{2H}+g(t)\right).
\label{generalizedcontinuity}
\end{equation}

As argued in \cite{Komatsu2013b}, the assumption of a non-adiabatic-like expansion of the universe simplifies the model by considering a dependence of the form $f(t)=\alpha H(t)^2$. By comparing Eq. (\ref{continuityeq}) with (\ref{generalizedcontinuity}), and Eq. (\ref{accelerationeq}) with (\ref{generalizedacceleration}), we get $\alpha=0$. Finally, the Friedmann equation can be written as follows:

%\begin{equation}
%\left(\frac{\dot{a}}{a}\right)^2=\frac{8\pi G\rho}{3}+\alpha H^\beta,\;\;(\beta>0)
%\end{equation}

%\nd where we supposed that the extra entropic-force term also depends on a power of $H$. However, Eq. (\ref{accelerationeq}) and (\ref{continuityeq}), assuming a spatially flat universe, imply $\alpha=0$, and therefore

\begin{equation}
\left(\frac{\dot{a}}{a}\right)^2=\frac{8\pi G\rho}{3}.
\label{friedmanneq}
\end{equation}

The three main equations of the generic entropic-force model are (\ref{accelerationeq}), (\ref{continuityeq}), and (\ref{friedmanneq}), but only two of them are independent.

%262+16x11=438
%+1501=1939 en total

%\subsection{Solution of the model}

We obtain the solution of the model under the assumption of a homogeneous, isotropic, and spatially flat universe. This solution describes the evolution of the Hubble parameter $H$ with the scale factor $a$. From Eq. (\ref{accelerationeq}), (\ref{continuityeq}), and (\ref{friedmanneq}), we obtain

\begin{equation}
\frac{H}{H_0}=\left(\frac{a}{a_0}\right)^{\frac{2\,C_d-3(1+\omega)}{2}},
\label{solutionnu}
\end{equation}

\nd where $\omega=\frac{p}{\rho\,c^2}$, $a_0$ and $H_0$ being the contemporary values of $a$ and $H$, respectively. A straightforward calculation (first-order ordinary differential equation) yields the following explicit time-dependent solution:

\begin{equation}
\frac{a}{a_0}=\left[\frac{3+3\omega-2C_d}{2}H_0(t-t_0)+1\right]^\frac{2}{3+3\,\omega-2\,C_d}.
\end{equation}

Let us focus now on the simple case of non-relativistic matter-dominated universe, i.e. $\omega=0$. The deceleration parameter $q\equiv-\ddot a/(aH^2)$ is then given by the following constant:

\begin{equation}
q=-\frac{1}{2}(2C_d-3)-1.
\label{deceleration}
\end{equation}

Values of $q<0$ correspond to an accelerating universe and $q>0$ to a decelerating one. The deceleration parameter does not depend on time (neither on $a$ nor on $H$) and, therefore, it is unable to explain periods of acceleration and deceleration. This clearly contradicts the well established fact that a matter-dominated phase, with $q=+0.5$, is necessary for structures to form while the accelerated expansion emerges from the transition to negative $q$ values at late times. As already noticed in \cite{Komatsu2013,Perico2013,Komatsu2013b,Basilakos2012}, a viable cosmology can not be fully accommodated within this oversimplified scenario. In the particular case of the Bekenstein-Hawking entropy (thermodynamically admissible if $d=2$), this was solved by considering correction terms in the scaling of the entropy \cite{Easson2012}. The inclusion of a first-order correction to the horizon entropy provides a natural source of inflation (accelerated expansion) of the early universe. Such correction is possible and necessary for $d \ne 2$. This is out of our present scope and constitutes the goal of an effort in progress. The aim of the present paper is to point out that some entropic-force models proposed in literature violate the thermodynamical Legendre structure. Providing a complete cosmological model that explains the different stages of accelerating and decelerating expansion constitutes the next step along this line.

%94+16x4=158
%+1939=2097 en total

\section{Comparison with supernova data}

\begin{figure}[h!]
\includegraphics[width=8.6cm]{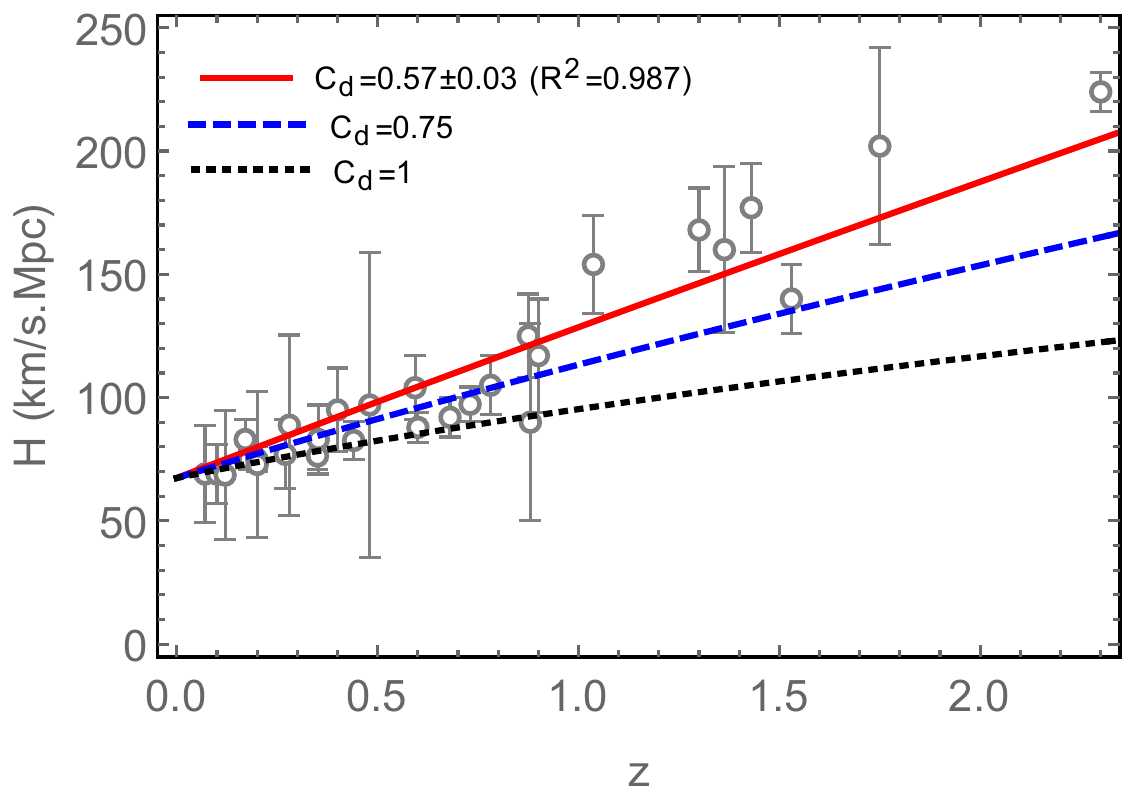}
\caption{\label{Hvsz} Hubble parameter $H$ versus redshift $z$. The open circle with bars are data points taken from Table 1 in \cite{Pradhan2021}. The black dotted curve is the EFS entropic-force model \cite{Easson2011} ($C_d=1$). The dashed blue curve is the entropic-force model with modified Hawking temperature \cite{Komatsu2013} ($C_d=0.75$). The solid red curve is a generic model, the optimal fitting giving $C_d=0.57\pm0.03$.}
\end{figure}

Supernova data are the mean source of available measurements in order to compare cosmological models. They constitute nowadays one of the best experimental tools for comparing various entropic-force models. We present here a simplified analysis of data, in order to determine a fitted value of the parameter $C_d$. In Fig. \ref{Hvsz}, we have plotted the Hubble parameter $H$ as a function of the redshift $z$ using the data points taken from Table 1 in \cite{Pradhan2021}. The equation describing $H(z)$ is obtained by replacing the definition of the redshift, $1+z\equiv a_0/a$, in Eq. (\ref{solutionnu}).

We have plotted three different entropic-force models. In all cases, the value of $H_0$ is set to be $67.4 km/s/Mpc$ based on the Planck 2018 results \cite{PlanckCollaboration2018}. The EFS model \cite{Easson2011} (black dotted curve) uses the Bekenstein-Hawking entropy and Hawking temperature; and therefore, $C_d$ is set to be $1$ ($A_d=2\pi$, $B_d=\pi$, and $d=2$). In the figure, we present also a particular -thermodynamically consistent- case, using $C_d=0.75$ (blue dashed curve). In the latter, $C_d$ is set equal to 0.75 because, according to \cite{Komatsu2013b}, this value was consistent with the supernova data observed at that time. This corresponds to a model using the Bekenstein-Hawking entropy and a modified Hawking temperature like in the KK model \cite{Komatsu2013}, i.e., $T=\gamma T_{BH}$. Finally, we show the generalized entropic-force model (solid red curve), where $C_d$ is determined by optimally fitting the data points. Of course, the red curve agrees better with the data, since the corresponding value of $C_d$ has been determined through fitting. This value constrains the relationship $C_d=d.A_d/B_d$, but it is not enough for determining an unique value for $d$.

\begin{figure}[h!]
\includegraphics[width=8.6cm]{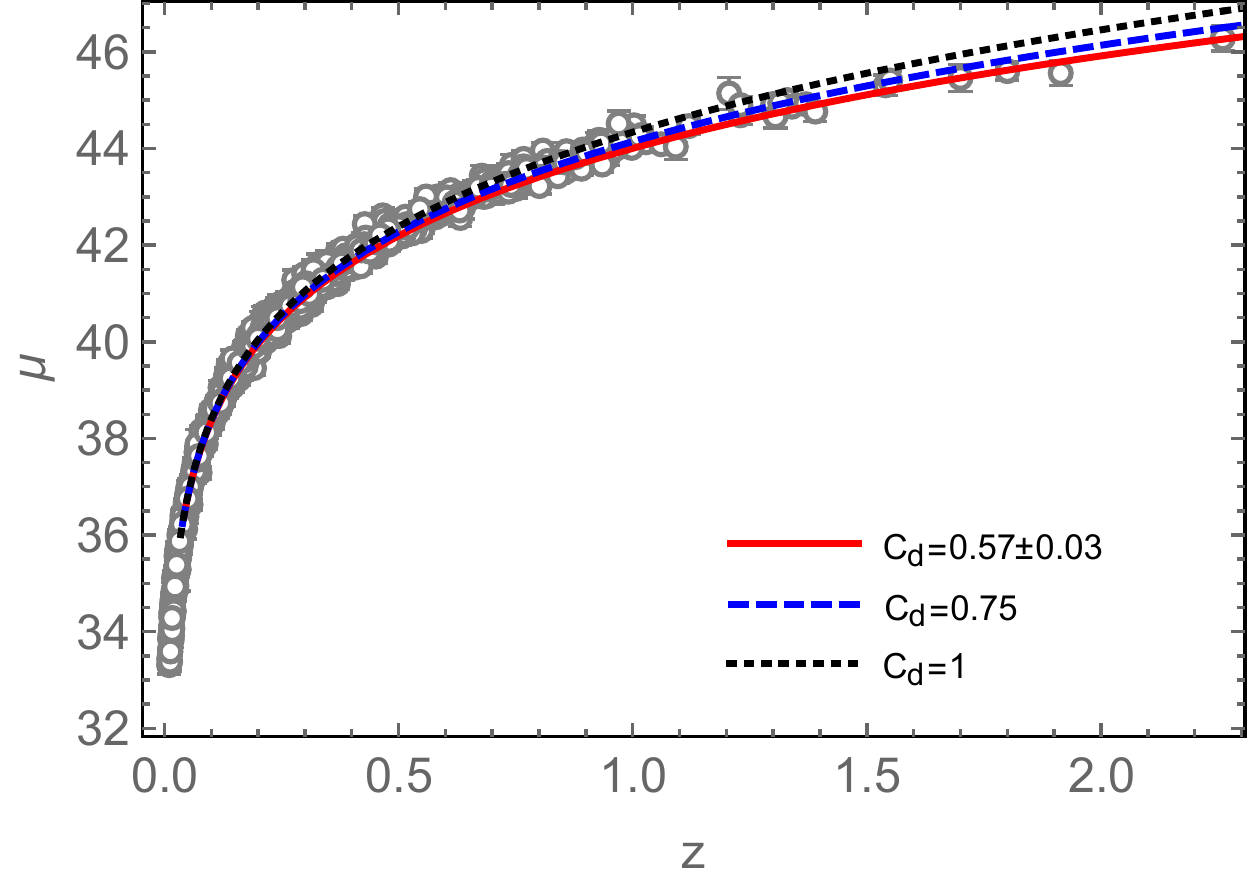}
\caption{\label{dL2} Dependence of the distance modulus $\mu$ with redshift $z$. The open circles with error bars are supernova data points taken from \cite{Scolnic2018}. SN1a absolute magnitude $M_0=-19.36$. The black dotted curve is the EFS entropic-force model \cite{Easson2011} ($C_d=1$). The dashed blue curve is the entropic-force model with modified Hawking temperature \cite{Komatsu2013} ($C_d=0.75$). The solid red curve is the model setting $C_d=0.57\pm0.03$. In all cases, $H_0=67.4$(km/s)/Mpc.}
\end{figure}

The previous data points exhibit in a transparent manner the consequences of different values of $C_d$. This is welcome since the distance modulus $\mu$ is not very sensitive to differing values of $C_d$. This lack of sensitivity can be appreciated in Fig. \ref{dL2}. The luminosity distance is an important parameter for investigating the accelerated expansion of the universe, and it is defined (see \cite{Komatsu2013,Easson2011} for instance) by

\begin{equation}
d_L(z)\equiv\frac{c(1+z)}{H_0}\int_1^{1+z}\frac{dy}{F(y)},
\end{equation}

\nd where $y\equiv a_0/a$, and $F(y)\equiv H(y)/H_0$. We remind that $\omega=0$. From Eq. (\ref{solutionnu}), we obtain

\begin{equation}
\begin{split}
&\frac{H_0}{c}d_L=\frac{2(1+z)}{2C_d-1}\left[(1+z)^{\frac{2 C_d-1}{2}}-1\right].
\label{dLnu}
\end{split}
\end{equation}
Notice that the luminosity distance depends indirectly on the dimension through $C_d$. In Fig. \ref{dL2}, we plotted the distance modulus $\mu$ versus redshift $z$ data taken from the so-called "Pantheon Survey", consisting of a total of 1048 SNe Ia \cite{Scolnic2018}, where% \cite{Riess2004} and \cite{Suzuki2012a}, 
\begin{equation}
\mu=5\log_{10}d_L-5,
\end{equation}

\nd with $d_L$ in parsec. Fig. \ref{dL2} displays the Pantheon Survey as the standard Hubble diagram of SN1a (absolute magnitude $M_0=-19.36$).
 
Also, we remark that the optimal fitting value of $C_d$ from Fig. \ref{dL2} is $C_d=0.50\pm0.02$ with $R^2=0.99998$. This is not shown in Fig. \ref{dL2} because it is visually indistinguishable from the red solid curve.

%We realize that the luminosity distance fitting is not very sensitive to different values of $C_d$ or even to different models. Indeed, all three entropic-force models agree well with the modulus distance data points. In fact, there exist numerous and diverse cosmological models with a good agreement with such data (see, for example, \cite{Komatsu2014,Tartaglia2009,Knop2003,Jana2014,Zhang2008}).

%511+16x3+120x3=919
%+2097=3016 en total

\section{Discussion and conclusions}

In conclusion, we analyzed thermodynamically admissible models based on entropic-forces. This approach provides, in contrast with the dark energy description, a concrete physical understanding of the acceleration. The accelerated expansion rate is the inevitable consequence of the entropy associated with the information storage in the universe.

In order to examine the entropic cosmology, we have introduced extra terms from a generalized entropy in the cosmological equations, assuming that the horizon of the universe has associated entropy and temperature. The main contribution of this paper is to show that the independent choice of the entropy and temperature of the horizon may violate the Legendre structure of thermodynamics. This is the case, for example, of the KK model discussed in \cite{Komatsu2013}, whereas the first EFS entropic-force model \cite{Easson2011} is consistent with thermodynamics. The way of avoiding the inconsistencies is to adapt the temperature to an extensive entropy. Consequently, the $H^2$ entropic-force term is derived from a generalized entropy, similarly to the original entropic-force model \cite{Easson2011}. It is on this basis that we have formulated the modified Friedmann, acceleration, and continuity equations. We show that the Friedmann equation itself does not include the entropic-force term, in variance with the continuity and acceleration equations.

We have obtained a solution of the model, assuming a homogeneous, isotropic, and spatially flat universe. We have confirmed that entropic-force models constitute a plausible alternative to explain an expanding universe. However, the simplest versions of these models cannot describe correctly the periods of acceleration and deceleration, since the entropic-force term is in all cases of the $H^2$-type. Easson, Frampton, and Smoot have proposed a way of overcoming this difficulty by including a subextensive correction term in the scaling of the entropy \cite{Easson2012}, specifically a logarithmic term. The discussion of a cosmologically more satisfactory model which contains the EFS model as a particular case is in progress. Let us anticipate that the addition of thermodynamically subdominant terms does not modify the basic Legendre transformation structure, which only depends on the dominant term.

Finally, we compared the performance of the entropic-force model with different values of the parameter $C_d$ with regard to the recently available supernova data. This allows us to identify the best value of $C_d$. Fitting the distance modulus gives us the optimal value $C_d=0.50\pm0.02$, while the Hubble parameter $H$ as a function of the redshift $z$ gives us $C_d=0.57\pm0.03$. These values are clearly different from that of the first EFS model ($C_d=1$) and from that obtained by Komatsu and Kimura in \cite{Komatsu2013b} ($C_d=0.75$). This imposes a constraint to the relationship $C_d=d.A_d/B_d$, but it does not suffice for determining a value for the dimension $d$. Indeed, we remind that $A_d$ and $B_d$ are the factors appearing in the entropy and the temperature equations, respectively. These values imply that the temperature of the Hubble horizon differs from the usually assumed Hawking temperature.

As a serious alternative to mainstream cosmology, entropic-force models need to satisfactory handle three important points: (i) validation through the full data analysis, including the covariance matrix; (ii) correct explanation of the different periods of acceleration and deceleration; and (iii) a physical principle that mandates the entropy and temperature to be used for the Hubble horizon. In the present paper we have focused on the last point.

\section*{Acknowledgments}
One of us (C.T.) acknowledges fruitful discussions with G. 't Hooft. 
This work has received financial support from CAPES, CNPq, and FAPERJ (Brazilian agencies).

%\appendix

%\bibliography{entropiccosmology}

%2681 words+27eq x16+2fig x120=3353

\end{document}